\documentclass[12pt,preprint]{aastex}
\begin{document}
\normalsize

\slugcomment{Submitted to ApJ, November 6, 2006}
\slugcomment{Revised, July 15, 2007}

\shorttitle{WMAP and HI}
\shortauthors{Verschuur}

\title{High Galactic Latitude Interstellar Neutral Hydrogen Structure and Associated (WMAP) High Frequency Continuum Emission }

\author{Gerrit L. Verschuur}
\affil{Physics Department, University of Memphis,
Memphis, TN 38152
gverschr@memphis.edu}

\begin{abstract}
Spatial associations have been found between interstellar neutral hydrogen (HI) emission morphology and small-scale structure observed by the {\it Wilkinson Microwave Anisotropy Probe (WMAP)} in an area bounded by {\it l} = 60\arcdeg\  \& 180\arcdeg, {\it b} = 30\arcdeg\  \& 70\arcdeg, which was the primary target for this study.  This area is marked by the presence of highly disturbed local HI and a preponderance of intermediate- and high-velocity gas.  The HI distribution toward the brightest peaks in the {\it WMAP} Internal Linear Combination ({\it ILC}) map for this area is examined and by comparing with a second area on the sky it is demonstrated that the associations do not appear to be the result of chance coincidence.  Close examination of several of the associations reveals important new properties of diffuse interstellar neutral hydrogen structure.  In the case of high-velocity cloud MI, the HI and {\it WMAP ILC} morphologies are similar and an excess of soft X-ray emission and H${\alpha}$ emission have been reported for this feature.  It is suggested that the small angular-scale, high frequency continuum emission observed by {\it WMAP} may be produced at the surfaces of HI features interacting one another, or at the interface between moving HI structures and regions of enhanced plasma density in the surrounding interstellar medium.  It is possible that dust grains play a role in producing the emission.  However, the primary purpose of this report is to draw attention to these apparent associations without offering an unambiguous explanation as to the relevant emission mechanism(s).   
\end{abstract}


\section{Introduction}

Examination of the {\it Wilkinson Microwave Anisotropy Probe (WMAP)} images immediately fires the imagination, especially the summary image that has received so much publicity, the so-called {\it Internal Linear Combination (ILC)} map (Hinshaw et al, 2006).  It is readily accessed at http://map.gsfc.nasa.gov and was produced after subtraction of suspected galactic components from the 5-channel observations carried out by {\it WMAP} and then combining the data.   But do the remaining structures in the {\it ILC} map truly reveal the fingerprints of processes that took place shortly after the universe was born?  Upon close inspection, certain features in the {\it WMAP ILC} map (hereafter the {\it ILC} map) look hauntingly familiar to those who have spent their careers studying Galactic, interstellar neutral hydrogen (HI) structure.  This can be recognized in a visual comparison of the above-referenced {\it ILC} image and the all-sky HI column density map found at http://www.astro.uni-bonn.de.  Several extended areas of excess emisison at high galactic latitudes ({\it b} $>$30\arcdeg) are present in both maps; for example in the areas around galactic longitude and latitude ({\it l,b})  = (170\arcdeg,80\arcdeg), which exhibits extensive intermediate- and high-velocity HI, and around ({\it l,b}) = (160\arcdeg,-35\arcdeg).  Also, a striking area of extended {\it ILC} structure that reaches from ({\it l,b}) = (230\arcdeg,0\arcdeg) through ({\it l,b}) = (315\arcdeg,-35\arcdeg) has a counterpart in the HI data where a tongue of emission follows roughly the same axis, emerging from the galactic disk HI at ({\it l,b}) = (255\arcdeg,-15\arcdeg) and stretching through ({\it l,b}) = (310\arcdeg,-35\arcdeg).  These apparent associations initiated a closer look at the HI structure and {\it ILC} data for several areas of sky.  If the {\it ILC} small-scale structures correspond to cosmological signals, absolutely no associations with HI structure, other than those due to pure chance, should be found.     

The thrust of this paper is to highlight the fact that a close relationship between {\it ILC} and HI structures appears evident in the data.  This deserves closer attention.  Lagache (2003) has examined the relationship between {\it WMAP}  structure and interstellar HI by considering the HI emission integrated over all velocities as a function of position and suggests that excess {\it WMAP} emission may be associated with small, transiently heated, dust particles.  In the data presented below, the apparent associations are found by examining HI area maps produced by integrating over limited velocity ranges as opposed to the total HI content.  Small-scale HI features due to relatively weak and narrow emission profile structure tend to become lost when maps of integrated HI content are plotted.     

In \S 2 the data and analysis are described.  In \S3 several of the most informative examples of close correspondences and morphological associations between HI and {\it ILC} structure are presented.  In the discussion in \S4 reference is made to possible, dust-related emission mechanisms that may be invoked to account for the associations and a geometric model for producing the apparent morphological associations is outlined.  In the conclusions in \S5 the hope is expressed that others will be motivated to seek confirmation (or otherwise) of the claims made here.  

\section{Data and Analysis}
The goal of the present study was to determine if parallels exist between small-angular-scale (1\arcdeg\ to 2\arcdeg) galactic HI structure and high-radio-frequency continuum emission features observed by {\it WMAP}.  Neutral hydrogen emission profile data with an angular resolution of 0.6\arcdeg\ were obtained from the Leiden-Dwingeloo, sidelobe-corrected HI Survey (Hartmann \& Burton, 1997) as well as the more extensive Leiden-Argentina-Bonn ({\it LAB}) All-Sky Survey (Kalberla et al.  2005).  In our study the distribution of the HI brightness was initially plotted as a function of galactic longitude and latitude ({\it l,b} maps) every 10 km/s in velocity integrating over a 10 km/s range.  The focus of the first phase of this study is Region A, bounded by {\it l} = 60\arcdeg\ \& 180\arcdeg, {\it b} = 40\arcdeg\ \& 70\arcdeg\ for which 25 maps covering the velocity range of -200 to +50 km/s with respect to the local-standard-of-rest were produced. 

The HI and  {\it ILC} data in the form of {\it l,b} maps were visually examined in a manner akin to blink comparison using a transparent overlay of the  {\it ILC}  data to search for apparent associations between small--scale structures.  The galactic HI sky is filled with small-scale structure evident in such {\it l,b} maps (e.g., see Hartmann \& Burton, 1997) so that any attempt to compare with small-scale structure found in the {\it ILC} data, which also consists of many small peaks of order 1 to 2\arcdeg\ across, will lead to chance agreements in position.  This was tested (to first-order) by also comparing the {\it ILC} map for the Region A data with the HI {\it l,b} maps for Region B, chosen as {\it l} = 180\arcdeg\ to 300\arcdeg, {\it b} = 40\arcdeg\ to 70\arcdeg.  This is located symmetrically opposite the galactic anti-center with respect to the Region A.  To this end, 20 HI maps from -120 to +80 km/s at 10 km/s intervals were produced to cover the extent of the galactic HI emission in Region B.  

For Regions A \& B, lists of the brightest {\it ILC} peaks were produced.  For Region A this included all 51 {\it ILC} peaks with amplitudes $\geqq$0.100 mK.  For Region B, 51 peaks were also used whose amplitudes are $\geqq$0.117 mK.  The centers and amplitudes of the  {\it ILC} and the apparently associated HI peaks were recorded and the angular separations between the {\it ILC} and apparently associated HI peaks calculated.  The same was done for two lists of 51 negative amplitude {\it ILC} peaks for Regions A \& B whose amplitude limits are $\leqq$-0.131 \& $\leqq$-0.144 mK respectively.      

\subsection{Statistics of the Apparent Associations}
The morphologies of both HI and {\it ILC} structures are complex yet distinct brightness features of a small angular extent of order 1\arcdeg\ to 2\arcdeg\ are common in both classes of data.  To derive a rough estimate of the likelihood of finding chance associations between {\it ILC} and HI peaks it is possible to calculate how likely it is that one of N$_{Wpeaks}$ in the {\it ILC} map will coincide by chance with one of N$_{Hpeaks}$ in the HI maps for the the same area, A, assuming both to be randomly distributed on the sky.  

The number, N, of HI peaks expected to lie by chance within a distance r of any {\it ILC} peak, that is within an area of order ${\pi}$ r$^{2}$ square degrees centered on the {\it ILC} peak, is given by
\begin{equation}
N ={\pi} r^{2}  A^{-1} N_{Wpeaks}\  N_{Hpeaks}.
\end{equation}

This is a first-order approach to the problem and ignores an edge-effect that will be introduced for large separations, r, which reduces the effective area, A, to be entered in Eqtn. 1.  At small separations this is a minor correction.  This estimate is in any case limited for several other reasons.  For example, the two classes of structure are treated as being random on the sky, which is not so for HI, because it is known to be filamentary.  Also, the apparent associations that are found (see Table 1 below) are spread over all velocities with associations at anomalous velocities relatively over-represented in Region A.  In addition, no account is taken of morphological similarities between the two types of structure other than to search for sets of closed contours on a scale of 1\arcdeg\ to 2\arcdeg.  Morphological similarities reported in \S3.3 below are difficult to quantify in a formal manner but in any event would reduce the likelihood of chance associations. 

For each of the HI maps about 10 to 15 contour levels were plotted and the number of peaks, defined as sets of closed contours with dimensions of order 1\arcdeg\ to 2\arcdeg, was counted on each of the HI maps.  A given HI peak on one contour map can usually be followed over at least two or three adjacent maps for HI at velocities outside $\pm$\ 20 km/s with respect to the local-standard-of-rest.  In one case the relevant HI peak could be followed over as many as 8 maps (i.e., 80 km/s).  Furthermore, low velocity HI ($-20<v<20$\ km/s), narrow emission line components of order 6 km/s wide tend to appear in only one contour map integrating over a 10 km/s width. On average, each HI peak is identified in 2 adjacent velocity maps so that the value of N$_{Hpeaks}$ is this half of the number found in all the HI {\it l,b} maps used in the analysis.   

 In the case of Region A, about 452 obvious HI peaks in the 25 area maps were counted, which sets N$_{Hpeaks}$ = 226.  For Region B the HI is more highly concentrated to low latitudes and low velocities and fewer HI peaks were counted in the 10 km/s wide HI maps at b$\geqq$40\arcdeg.  Using the same criterion that on average an HI peak can be followed over 2 adjacent HI maps, N$_{Hpeaks}$ = 130 is used in the calculations. 

\subsection{The Angular Distribution of Apparent Associations}
In Fig. 1 the distribution of the angular offsets in 0.\arcdeg2 bins between {\it ILC} and HI features for a number of data comparisons are shown.  Fig. 1(a) presents the histogram of angular separations between {\it ILC} and HI peaks when the {\it ILC} peaks for Region A are compared to the HI data for Region A.  It shows a clear excess at small angular separations with the peak occurring at about 0.\arcdeg8, only slightly larger than the beamwidth.  Beyond 1\arcdeg\ separation, the values fall below expectations for chance associations indicated by a dashed line that represents Eqtn. 1.  Of the 51 {\it ILC} peaks in Region A, some 60 associations make up the histogram.  As will be discussed below, this carries a clue to the nature of the phenomenon producing the associations.  

A similar excess of offsets between {\it ILC} and HI peaks with small angular separations is seen in Fig. 1(b) for Region B, which shows the results obtained by overlaying the {\it ILC} data on the HI maps for the same area.  The numbers involved are smaller than for Region A with the bulk of the HI concentrated to low velocities.  
 
An estimate of the significance of the apparent associations found in Figs. 1(a) and (b) can be obtained by overlaying the {\it ILC} data for Region B on the HI peaks for Region A, and vice versa.  Figure 1(c) shows what is found when the Region B {\it ILC} peaks are overlain on Region A HI maps.  Figure 1(d) shows the converse.  Neither plot shows evidence for systematic small-scale angular offsets other than might be expected from chance.  

Finally, negative {\it ILC} peaks and HI structure were compared.  Figure 1(e) shows the results found by overlaying Region A negative amplitude peaks on Region A HI maps and Fig. 1(d) shows the results derived by overlaying Region B negative {\it ILC} peaks on HI data for the Region B.   Fig. 1(g) plots the results obtained when comparing the negative {\it ILC} peaks for Region B with the HI for Region A and Fig. 1 (h) shows the converse.  None of these four plot show convincing evidence for systematic small-scale angular offsets other than might be expected from chance.

Another way to summarize these results is to compare the total number of cases of close associations at small angular separations.  For example, for those offset by 1.\arcdeg1 or less, Region A contains 41 cases, compared to the 23 estimated from chance according to Eqtn. (1), which compares to 20, 17, 21 cases for the plots in Figs. 1 (c),( e) \& (g), which are independent estimates of what is to be expected due to chance associations.  Similarly, for Region B the direct comparison produces 22 cases of small offsets compared to 13 predicted by chance, with the other three comparisons summarized in Figs. 1 (d), (f)  \&(h) producing 10, 9 \& 6 cases respectively.  This suggests that the close associations found in Region B are also significant.  

These plots point to the possibility that the {\it ILC} continuum peaks are located at the boundaries of galactic HI emission features, which implies at the interface between HI structures and surrounding plasma, or at the interface between colliding HI structures.  

\section{Case Studies of Associated HI and Continuum Structures}
If the close associations between {\it ILC} and HI structures found especially in Region A are indeed significant then they should be expected to reveal underlying aspects of interstellar gas dynamics that may allow the cause of the relationships to be understood.  That would then remove the study from the realm of a statistics to that of interstellar physics.

The cases outlined below emerged from the study of an extended version of Region A with the latitude boundary set to 30\arcdeg.  Table 1 lists the apparent associations with an identifying number for the {\it ILC} peak given in column 1 and its galactic coordinates (in degrees) in columns 2 \& 3.  The peak {\it ILC} brightness temperature in mK is given in column 4.  Column 5 gives the center velocity in km/s with respect to the local standard of rest of the HI map used to determine the positions of the relevant HI feature; hence -95 km/s refers to the HI map integrating from -100 to -90 km/s.  Columns 6 and 7 give the galactic coordinates for the HI peak with the amplitude in units of K.km/s in Column 8.  The angular offset in degrees between associated peaks is given in Column 9.  

Table 1 includes 64 {\it ILC} peaks with amplitudes $\geqq$0.100 mK.   All but two of these appear to have associated HI peaks yet the table includes 83 entries for HI associated with the remaining 62 {\it ILC} peaks.  This is consistent with multiple HI structures being involved in creating some of the continuum emission peaks.  

In those directions where the most dramatic associations between {\it ILC} and HI structure were noted, the HI data were examined in more detail by using HI maps made by integrating over 5, 2 or 1 km/s and latitude-velocity {\it b,v } plots with a view to obtaining further insights into the nature of the possible relationship between the two forms of emission. 

In most of the plots below, the {\it ILC} data are shown as contours overlain on the HI morphology plotted using inverted gray-scale shading and these examples are offered because they each reveal something interesting and unexpected about the nature of interstellar matter.  This is regarded as highly significant because these directions would never have been chosen for closer study if it were not for the apparent associations with the continuum emission peaks highlighted in the {\it ILC} map.  

\subsection{Source 25 at ({\it l,b}) = (112.\arcdeg3,57\arcdeg.8): Directly overlapping HI and continuum features} 
Figure 2(a) shows the brightness of a high-velocity (v ${\leqq}$-100 km/s) HI feature \# 25 from Table 1 centered at -118 km/s integrated over 5 km/s (peak value 51 K.km/s).  Fig. 2(b) shows the brightness of an intermediate-velocity (between -100 and -30 km/s) HI feature centered at -87 km/s integrated over 5 km/s (peak value 47 K.km/s).  By using Gaussian analysis of the HI profiles every 0.\arcdeg 5 in latitude and longitude across the peaks, the coordinates of the centers of the HI structures were accurately derived and found to be identical in latitude (to $<$ 0.\arcdeg 1) and also identical to the latitude center of the {\it ILC} peak.  The centers of two HI features are identical in longitude but offset from the {\it ILC} peak by 0.\arcdeg 3, half the the beamwidth used in the HI studies. 

Examination of the latitude-velocity {\it b,v} plot at {\it l} = 112\arcdeg\ for this feature revealed no significant HI emission at -100 km/s between the two peaks seen in Figs. 2(a) and (b).  However, a marked lack of low-velocity HI emission around zero velocity is revealed.  (Low-velocity is defined as between + 30 and -30 km/s.)  Fig. 2(c) shows the integrated HI content over the velocity interval -8 to +2 km/s with the continuum contours overlain.  The peak in the {\it ILC} structure is clearly co-located with a lack of low-velocity HI. Also at low velocities an HI peak at ({\it l,b}) = (115\arcdeg , 57\arcdeg) is associated with a secondary peak in the continuum emission.  Further relationships emerge when the integrated HI emission between -130 \& -120 km/s is examined, Fig. 2(d). Two HI maximum at ({\it l,b}) = (113\arcdeg,55\arcdeg) and ({\it l,b}) = (115\arcdeg,54\arcdeg) overlie HI minima seen in Fig. 2(c).  

These plots suggest a direct relationship between high- and intermediate-velocity HI and a distinct minimum in low-velocity HI, with all of them related to the presence of an {\it ILC} continuum emission peak.  This contrasts with those models that would place the high-and intermediate-velocity HI at very different distances well removed from local HI.  For example, Wakker (2001) places the high-velocity gas at distances of several kiloparsecs with the intermediate-velocity HI at about 1 kpc, both of which contrast with the realm of high-latitude, low-velocity H, which is local at distances of order 50 to 100 pc.  Blaauw \& Tolbert (1966) originally noted that intermediate-velocity HI and a relative lack of low-velocity gas is a hallmark of that area of sky encompassed by Region A, which places them both within about 100 pc of the Sun.  Here we find striking evidence in Fig. 2 that HI structure in all three velocity regimes is related, something that would not have been noticed but for the apparent association with a significant peak in the {\it ILC} structure.  This implies that HI gas in all three velocity regimes in this direction is local.  (The possibility that small-scale, intermediate-velocity structure is associated with a lack of low-velocity gas has been briefly noted by Burton et al. (1992) as well as Kuntz \& Danly (1996) without their following up on the implications.) 

\subsection{Sources 11 \& 31 centered at ({\it l,b}) = (119.\arcdeg5, 57\arcdeg) and association between high- and low-velocity HI}
Figure 3(a) shows the HI structure of high-velocity gas integrated between -140 and -110 km/s associated with two {\it ILC} continuum peaks, 11 \& 31 (Table 1) identified in Fig. 3(b).  The {\it ILC} peaks are linked by a ridge of emission straddled by two HI features (peak amplitudes 22 K.km/s) whose morphologies closely follow the continuum radiation contour lines.  Gaussian analysis of the HI profiles in a 0.\arcdeg5 grid for the entire area covered in Fig. 3(a) was carried out and allowed the center velocities of the two HI components to be determined.  They are -127 and -118 km/s for the northern and southern peaks, respectively.  The Gaussian analysis also allowed the HI column density for two other features to be mapped, shown in Figs. 3 (b) and (c).  Their presence was recognized in a set of HI emission profiles at {\it l}=120\arcdeg\ shown in Fig. 3(d).  Most striking is a component at -8 km/s, that is seen only at ({\it l,b}) = (120\arcdeg,57\arcdeg).   It has a peak column density of 7$\times$10$^{18}$ cm$^{-2}$ and its morphology is plotted in Fig 3(b), which shows that it is located precisely on the saddle in the continuum contours between two peaks.  Even more striking is the fact that it is unresolved in angle.  

Fig. 3(c) shows the column density plot for the -17 km/s feature evident in Fig. 3(d), again based on the results of the Gaussian analysis of the area profiles (peak column density 19$\times$10$^{18}$ cm$^{-2}$).  The location of this component also appears related to the presence of the continuum ridge with the southern peak located at the position of the HV peak that nestles in the indentation (or pinch) in the continuum emission ridge.  At this stage of the analysis, no clear relationship between intermediate-velocity HI structure and other features in the area of Fig. 3 has been noted.  However, very weak positive velocity HI emission is also found toward this structure but it will require confirming observations to determine if it is real. 

The plots in Fig. 3 represent the second example of a close relationship between HI structures at low- and high-velocities, with each of them related to the presence of an {\it ILC} feature.  In addition, the discovery of the angularly unresolved, low-velocity, high-galactic-latitude emission peak is unprecedented and deserving of observations at much higher angular resolution.

\subsection{Continuum source 33 at (l,b) = (89.3\arcdeg, 34.5\arcdeg) and associated high-velocity HI}
Figure 4 illustrates the dramatic association between {\it ILC} emission peak 33 (Table 1) and HI.  Here nine inverted gray-scale images of the HI at different velocities are overlain by a contour map of the {\it ILC} structure. All the HI plots are single channel maps (1 km/s wide) and the peak brightness temperatures are of order 0.8 K, except for the map at -200 km/s where the peak is 0.3 K and at -150 km/s where it is 1.1 K.  The limits of the associated HI emission are -135 km/s at the same location as the peak at -140 km/s, and -205 km/s.  The HI centroid of emission shifts in velocity as it follows the ridge of the {\it ILC} emission feature with a bifurcation of the HI peak starting at -170 km/s to produce two components that "move" along opposite sides of the continuum ridge as the velocity is increased.  

A preliminary attempt to sketch the structure revealed in these and HI {\it l,b} maps at smaller velocity intervals suggests a twisted pattern around an axis defined by the continuum radiation, possibly related to helical magnetic field structure around this axis.  Higher resolution HI data are desirable to untangle the HI structure in this fascinating area.  Note that the map of total HI in these directions carries no hint of the existence of the structure seen in Fig. 4 because it is very faint compared to low-velocity gas in this area of sky.

Examination of {\it b,v} plots and emission profile data for this area reveal further details that are relevant to understanding the apparent relationship between HI and {\it ILC} features.  This is illustrated in Fig. 5(a) in which several HI emission profiles cutting across the continuum feature seen in Fig. 4 are shown.  In a manner similar to that reported in \S3.2 above, a distinct additional component, in this case at -40 km/s, emerges in one of the profiles.  Its morphology integrated between -42 to -38 km/s  is shown in Fig. 5(b) as an inverted gray scale image (peak value 6.7 K.km/s) with the same {\it ILC} contours shown in Fig. 4 overlain.  Fig. 5(c) illustrates in contour map form how closely this intermediate-velocity HI structure (contours from 3 K.km/s in steps of 0.05 K.km/s) mimics the continuum contours seen in (b).  Their axes are aligned to better than 5\arcdeg.  This HI component is clearly associated with the continuum as well as the HI structure at high velocities seen in Fig. 4.  

Further examination of the {\it b,v} contour maps reveals a phenomenon also found for the HI-continuum association noted in \S4.1 (Fig. 2) above, a dearth of low-velocity emission where the anomalous velocity HI structure shows a peak.  This is illustrated in Fig 5(d) where the integrated HI emission from -5 to +15 km/s (peak 36 K.km/s) is shown.  A distinct minimum (36 K/km/s) compared to the maximum value in this plot (112 K.km/s) in the low velocity integrated HI emission is found at the location of the peak in the intermediate-veloicity HI emission at -40 km/s.  This, in turn, coincides with the location of continuum source 32 and the HV structure seen in Fig. 4.  

\subsection{Close-up view of HVC MI}
Figure 6 shows the inverted gray-scale image of the HI brightness associated with high-velocity cloud MI integrated over the velocity range -140 to -100 km/s.  The double HI peaks are offset from and parallel to a pair of {\it ILC} peaks, sources 19 and 57.  A third {\it ILC} peak 43 is also associated with a weak HI feature, see Table 1.  This main double HI feature is well known and corresponds to high-velocity cloud HVC MI.   

A crucial clue as to the likely cause for associations between the small-scale HI and high-frequency continuum structures is found in a report of excess soft X-ray emission toward HVC MI found by Herbstmeier et al. (1995).  Their Figure 7a has been adapted to correspond to the data in Figs. 6(a) \& (b) and is shown as Fig. 6(c) where the contours correspond to the HI emission from HVC MI similar to the data in (a).  The shaded pixels overlain correspond to areas of excess soft X-ray emission.  Furthermore, Tufte, Reynolds \& Haffner (1998) have reported excess H${\alpha}$ emission at several positions toward HVC MI.  The close spatial relationship seen in Fig. 6(c) between the X-ray hot spots and the HI and the high-frequency continuum emission observed by {\it WMAP}, as well as the presence of H${\alpha}$ emission contains critical clues that should lead to understanding the nature of the associations between HI and weak, high-frequency continuum emission.   It is predicted that high-velocity cloud MII, a companion to MI but located outside Region A, will show a similar relationship between HI and {\it ILC} structures.  (The outcome of this prediction will be discussed in a future report.)

In Fig. 6(d) the {\it ILC} contours are overlain on the low-velocity HI data.  Again a relative lack of LV emission, here integrated between Ð5 \& +5 km/s, is seen to be closely associated with the continuum and HI structure at high velocities.  The HI column density is 1.8 K.km/s for the minimum compared to 6.2 K.km/s for the maximum at the top left of (d).

\section{Discussion}
Several authors have considered the possibility that excess {\it WMAP}  emission may be produced by spinning dust grains, in turn possibly associated with HI.  For example, Davies et al. (2006) note that the spectrum of the low {\it WMAP}  frequency data is consistent with such a model.  In contrast, Banday et al. (2003) find evidence in earlier COBE data of a component in the continuum emission with a dust-like morphology but with a synchrotron-like spectrum.  Larson \& Wandelt (2004) have suggested that both the {\it ILC} data positive and negative peaks do not have sufficient amplitude to be accounted for by a cosmological interpretation of the data, which implies some other mechanism may be playing a role in generating the observed signals.  In a specific case involving the detection of anomalous microwave emission toward the Perseus molecular cloud, Watson et al. (2005) present an hypothesis that dipole emission from spinning dust grains can account for the spectrum of the continuum emission in the range 10 to 50 GHz.  An extensive literature exists concerning the possible role of spinning dust grains as the cause of continuum emission in the frequency range of the {\it WMAP}  experiment.  This includes Draine \& Lazarian (1998), Finkbeiner at al. (2002), Schlegel et al. (1998), and de Oliviera et al. (2005).  An alternative mechanism for producing low density electrons possibly capable of generating high-frequency radio emission involves a plasma physical phenomenon suggested by Verschuur (2007).

In general, many high latitude  {\it ILC} peaks listed in Table 1 for extended Region A have a corresponding, closely-spaced HI peak found in two or more {\it l,b} maps made at 10 km/s intervals.  The typical angular offset is approximately 0.\arcdeg8\ (Fig. 1a).  Including the cases below b = 40\arcdeg\ in the statistics makes no difference to the distribution seen in Fig. 1a.  Angular offsets of order 1\arcdeg\ between parallel HI, dust and H${\alpha}$ filaments have been reported by Verschuur et al. (1992).  

Based on the above results it is suggested that the {\it ILC} peaks are associated with HI structures that are interacting (probably colliding) with other HI structures, or interacting with regions of enhanced plasma density in surrounding interstellar space through which the HI is moving.  In one case illustrated above, two HI features at distinctly different velocities are coincident in position at the location of the {\it ILC} peak, Fig.2.  In other cases, two HI features at (nearly) the same velocity straddle a continuum emission peak in position, for example as shown in Fig. 3.  Depending on the geometry of the situation, positional coincidence will only be observed in those directions where a collision between two HI features (clouds) occurs along the line-of-sight.  However, where the HI features are interacting while moving along an axis oriented at some angle to the line-of-sight, the continuum emission peak will be observed as offset from the HI peak(s).  

A number of clues that may help account for the physics underlying the production of continuum emission at the surface of moving HI structures exist in the literature.  For example, as was described in \S3.4, Herbstmeier et al. (1995) report excess soft X-ray emission at the boundary of HVC MI where the {\it ILC} continuum structure is prominent (Fig. 6).  On a larger scale, Kerp et al. (1999) find evidence for widespread soft X-ray emission over much of Region A toward the high-velocity structures but the angular resolution of their data do not allow for closer comparison with our results.  Furthermore, Tufte et al. (1998) found H${\alpha}$ emission high-velocity clouds MI \& MII as well as other high-velocity features.  This raises tantalizing questions as to the emission mechanism that could produce the continuum radiation.  If it involves the formation of dust at the (shocked) interface between the HI and surroundings by a mechanism involving spinning dust grains such as has been proposed by Draine \& Lazarian (1998), the presence of both excess H${\alpha}$   and soft X-ray emission will also have to be taken into account. 

Lagache (2003) has searched for associations between HI and WMAP peaks but confined his study to the total HI content as a function of position.  In the examples above the amplitudes of the HI peaks show a relationship to {\it ILC} peaks that varies widely and in many cases, such as for the HI features shown in Figs. 4 or 5, are invisible in maps of total HI content over the relevant areas.   Land \& Slosar (2007) also studied the relationship between {\it WMAP} peaks and HI structure but considered only direct, point-to-point associations that are in fact relatively rare.  Far more likely are small angular offsets between peaks in the two forms of emission.  

Another interesting, possibly related phenomenon, has been found by Liu \& Zhang (2006) who studied the cross-correlation between {\it WMAP} and Egret ${\gamma}$ ray data and concluded that an unknown source of radiation, most likely of galactic origin, is implied by their analysis.  Such a source would produce foreground residuals that need to be removed in order to minimize their role in confusing the cosmological interpretation of the {\it WMAP} data.  Perhaps the source of this unknown radiation of galactic origin is to be found in processes occurring at the surfaces of Galactic HI structures moving through interstellar space and/or interacting with one another.  

Based on what is found in these examples, it is possible that where two HI features are interacting (colliding) current sheets are created in which particle acceleration may underlie the production of the continuum emission observed by {\it WMAP}.  Excess electrons at these interfaces may initially be introduced through the process described by Verschuur (2007) and references therein.  

To pursue this study this further, {\it ILC} structure should be compared with {\it l,b} HI maps integrated over smaller velocity ranges that the 10 km/s used in this overview and higher angular resolution HI observations of some of the most interesting HI features shown in Figs. 2 to 6 are desirable.  It is hoped that research into the relationships between interstellar HI morphology and high-frequency radio emission observed by {\it WMAP} will be stimulated by this work.

\section{Conclusions}
The goal of this study was to determine if evidence exists to suggest that small-scale structure in the {\it WMAP ILC} data and HI are related.  To that end, the map of {\it ILC} structure guided the study and led not only to the discovery of what appear to be highly significant spatial relationships between the two data sets but also drew attention to unexpected properties of the galactic HI that turned out to be especially interesting and relevant to accounting for the associations. 

In Region A clear associations have been found between small-scale structure observed by {\it WMAP} and galactic HI features identified in maps in which the HI is integrated over 10 km/s.  In Region B a similar excess of associated features was noted.  In contrast, no significant associations are found when comparing the {\it ILC} data for one Region with the HI data for the other Region, nor when the  minima in the {\it ILC} data are treated as peaks.  This argues that the apparent associations found in Region A are not due to chance.  While studying the data for Region B in which the HI is more concentrated to low velocities and latitudes it became clear that HI maps at smaller velocity intervals should be used in the search for associations.  A future report will describe such work together with extensive studies of structures at high latitudes in the southern galactic hemisphere.  
   
Taken as an ensemble, the spatial associations between HI and {\it ILC} emission peaks point to the existence of one or more processes occurring in interstellar space capable of generating weak continuum radiation observed by {\it WMAP}.  The radiation appears to originate at the surfaces of dynamic and interacting HI structures.  Where this interaction is viewed along the direction in which the HI is moving the radio continuum structure will overlap in position but where the HI has a transverse component of motion (much more common) the two forms of emission will appear closely offset on the sky.  

Of particular significance is the fact that the associations between {\it ILC} and HI structure discussed here led to the discovery that HI structures in the three velocity regimes (high, intermediate and low) are physically related to one another, which places them at a common distance from the Sun, probably of order 100 pc.
 
Rigorous studies of the apparent associations between  {\it ILC} and HI structures should make use of the full LAB survey spectral resolution of 1 km/s to produce HI maps integrating over 2 km/s velocity intervals, a limit set by what is known about low-velocity HI emission line structure.

\acknowledgments
I am grateful to Wayne Landsman and Gary Hinshaw for the first-year {\it ILC}\ continuum emission data in digital form and in rectangular coordinates.  The encouragement and advice offered by an anonymous referee as well as Richard Lieu as this paper evolved to completion was essential to its progress.  I am particularly grateful to Tom Dame for providing me with the necessary software for a Mac computer to allow the production of the vast majority of the final area maps of HI emission presented in this paper.  I also greatly appreciate encouragement from and discussions with Joan Schmelz  and Butler Burton and the important feedback offered by Dave Hogg, Mort Roberts, Ed Fomalont, Eric Priest, Petrus Martens, Tony Peratt, Max Bonamente, Shuang-Nan Zhang  \& Eric Feigelson as well as numerous audience members who attended seminars I gave in the early phases of this project.  
  
{}

\begin{deluxetable}{ccccccccc}
\tablecolumns{9}
\tablewidth{0pc}
\tablecaption{TARGET AREA ASSOCIATIONS}
\tablehead{
\colhead{No.} & \colhead{{\it l}}   & \colhead{{\it b}}    & \colhead{{\it ILC} Temp.} & \colhead{HI Velocity}    & \colhead{{\it l}}   & \colhead{{\it b}}    & \colhead{HI amplitude}  & \colhead{Angular offset } }
\startdata
(1) & (2) & (3) & (4) & (5) & (6) & (7) & (8) & (9)\\
1 & 139.6 & 40.1 & 0.227 & -55 & 139.2 & 40.5 & 37 & 0.50 \\
1 & 139.6 & 40.1 & 0.227 & 5 & 139.6 & 41.0 & 139 & 0.90\\
2 & 174.4 & 56.8 &0.223 &-95 & 175.0 & 55.6 & 8 & 1.24 \\
3 & 160.3 & 62.9 & 0.218 & -115 & 163.0 & 64.0 & 6 & 1.65\\
3 & 160.3 & 62.9 & 0.218 & -15 & 160.0 & 61.5 & 34 & 1.41\\
4 & 176.1 & 53.0 & 0.212 & 5 & 176.5 & 53.5 & 16 & 0.55\\
5 & 140.3 & 41.5 & 0.202 & -45 & 140.3 & 41.0 & 24 & 0.50 \\
6 & 172.3 & 59.5 & 0.199 & -65 & 168.2 & 58.5 & 27 & 2.31\\
7 & 174.0 & 50.5 & 0.182 & -125 & 171.5 & 51.4 & 4 & 1.83\\
7 & 174.0 & 50.5 & 0.182 & -15 & 173.1 & 50.1 &29 & 0.70\\
8 & 177.2 & 57.6 & 0.182 & -105 & 175.5 & 55.5 & 8 & 2.29\\
9 & 168.1 & 67.4 & 0.180 & -125 & 167.0 & 65.1 & 30 & 2.34 \\
9 & 168.1 & 67.4 & 0.180 & -25 & 168.2 & 66.5 & 26 & 0.90\\
9 & 168.1 & 67.4 & 0.180 & -45 & 168.3 & 68.5 & 44 & 1.10 \\
10 & 171.6 & 56.3 & 0.179 & -65 & 170.8 & 56.0 & 16 & 0.54 \\
10 & 171.6 & 56.3 & 0.179 & -55 & 172.0 & 56.0 & 13 & 0.37\\
11 & 118.1 & 57.0 & 0.175 & (-121) & 120.0 & 58.0 & 1.1 & 1.44\\
11 & 118.1 & 57.0 & 0.175 & (-115) & 120.0 & 56.5 & 1.0 & 1.15\\
11 & 118.1 & 57.0 & 0.175 & (-8) & 120.0 & 57.1 & 1.8 & 1.04\\
12 & 159.6 & 50.0 & 0.172 & none & none & none & \nodata & \nodata\\
13 & 169.5 & 46.5 & 0.172 & -15 & 167.5 & 47.6 & 42 & 1.76\\
13 & 169.5 & 46.5 & 0.172 & -5 & 170.0 & 47.2 & 47 & 0.78\\
14 & 174.7 & 48.2 & 0.172 & -15 & 175.5 & 46.5 & 28 & 1.78\\
15 & 174.4 & 31.8 & 0.172 & -5 & 174.0 & 31.0 & 110 & 0.87\\
16 & 101.8 & 59.5 & 0.169 & -5 & 100.8 & 58.5 & 32 & 1.12\\
17 & 78.2 & 55.9 & 0.168 & 5 & 78.5 & 56.0 & 62 & 0.20\\
17 & 78.2 & 55.9 & 0.168 & -25 & 79.5 & 56.5 & 26 & 0.94\\
18 & 82.3 & 55.7 & 0.168 & -15 & 80.0 & 56.7 & 35 & 1.64\\
19 & 165.7 & 64.8 & 0.168 & -125 & 167.0 & 65.0 & 30 & 0.59\\
19 & 165.7 & 64.8 & 0.168 & -15 & 165.9 & 65.4 & 18 & 0.61\\
20 & 91.5 & 36.7 & 0.166 & -135 & 91.0 & 38.1 & 15 & 1.46 \\
20 & 91.5 & 36.7 & 0.166 & -35 & 91.1 & 37.9 & 17 & 1.24 \\
20 & 91.5 & 36.7 & 0.166 & -15 & 91.9 & 37.0 & 36 & 0.44 \\
21 & 170.2 & 54.5 & 0.165 & -55 & 170.0 & 55.0 & 9 & 0.51\\
21 & 170.2 & 54.5 & 0.165 & 5 & 169.6 & 55.9 & 17 & 1.44\\
22 & 135.4 & 51.1 & 0.164 & -45 & 135.9 & 51.1 & 98 & 0.31\\
23 & 151.2 & 48.8 & 0.162 & -165 & 152.3 & 48.5 & 4 & 0.78\\
23 & 151.2 & 48.8 & 0.162 & -15 & 152.1 & 49.1 & 11 & 0.66\\
24 & 172.6 & 46.1 & 0.161 & -15 & 173.3 & 46.0 & 22 & 0.50\\
25 & 112.3 & 57.8 & 0.159 & (-87) & 112.8 & 57.6 & 4.8 & 0.33\\
25 & 112.3 & 57.8 & 0.159 & (-125) & 112.1 & 57.9 & 9 & 0.15\\
26 & 62.6 & 42.9 & 0.157 & -15 & 63.5 & 42.7 & 19 & 0.69\\
27 & 165.6 & 48.2 & 0.155 & -15 & 166.2 & 48.0 & 26 & 0.45\\
28 & 65.0 & 37.1 & 0.155 & -115 & 64.5 & 37.5 & 5 & 0.56\\
29 & 68.6 & 37.1 & 0.150 & -145 & 66.6 & 37.0 & 5 & 1.60\\
30 & 93.2 & 66.6 & 0.150 & -5 & 94.0 & 65.0 & 32 & 1.63\\
31 & 121.6 & 56.8 & 0.150 & -140 to -110 & see text & see text & see text & 1.00\\
32 & 113.4 & 66.4 & 0.149 & -15 & 115.4 & 66.0 & 22 & 0.90\\
33 & 89.3 & 34.5 & 0.146 & (-170) & 89.0 & 34.5 & 0.8 & 0.25\\
33 & 89.3 & 34.5 & 0.146 & -42 to -38 & 89.4 & 34.5 & 7 & 0.08\\
34 & 158.8 & 32.0 & 0.145 & -55 & 157.5 & 32.0& 30 & 1.10\\
35 & 94.6 & 38.6 & 0.143 & -165 & 94.7& 38.0 & 4 & 0.61\\
36 & 77.0 & 46.9 & 0.143 & -25 & 79.0 & 47.0 & 16 & 1.37\\
37 & 117.0 & 48.0 & 0.138 & -85 & 117.5 & 47.3 & 19 & 0.78\\
38 & 78.2 & 62.5 & 0.138 & -25 & 77.0 & 62.0 & 22 & 0.75\\
39 & 62.6 & 38.7 & 0.137 & -15 & 61.1& 39.5 & 22 & 1.42\\
40 & 167.3 & 55.7 & 0.136 & -135 & 167.0 & 55.5 & 16 & 0.26\\
41 & 172.8 & 35.9 & 0.132 & -5 & 172.0 & 36.0 & 25 & 0.66\\
42 & 174.7 & 63.6 & 0.131 & -55 & 172.7 & 65.0 & 32 & 1.66\\
43 & 163.8 & 63.8 & 0.130 & --125 & 162.1 & 63.6 & 4.5 & 0.78\\
44 & 131.8 & 57.6 & 0.128 & -75 & 131.0 & 57.5 & 30 & 0.44\\
44 & 131.8 & 57.6 & 0.128 & -115 & 133.0 & 57.0 & 33 & 0.44\\
45 & 125.2 & 40.0 & 0.127 & -65 & 123.9 & 40.5 & 23 & 1.11\\
46 & 138.7 & 52.7 & 0.126 & -55 & 139.3 & 52.5 & 16 & 0.41\\
47 & 166.3 & 34.6 & 0.125 & -45 & 165.8 & 35.0 & 36 & 0.57\\
48 & 118.5 & 68.3 & 0.124 & -25 & 119.0 & 68.0 & 19 & 0.35\\
49 & 132.9 & 50.0 & 0.122 & none & none & none & \nodata & \nodata\\
50 & 153.6 & 68.5 & 0.122 & -85 & 155.1 & 67.5 & 20 & 1.14\\
51 & 159.3 & 56.1 & 0.120 & -35 & 159.9 & 56.0 & 16 & 0.35\\
52 & 95.4 & 46.9 & 0.120 & -115 & 95.0 & 46.0 & 7 & 0.94 \\
52 & 95.4 & 46.9 & 0.120 & -65 & 96.0 & 47.5 & 7 & 0.73\\
53 & 149.4 & 67.2 & 0.118 & -5 & 149.6 & 67.5 & 56 & 0.31\\
54 & 160.7 & 57.4 & 0.115 & -85 & 160.0 & 58.0 & 5 & 0.71\\
55 & 142.7 & 35.2 & 0.115 & -25 & 143.7 & 34.5 & 12 & 1.20\\
56 & 167.3 & 61.6 & 0.114 & -5 & 167.9 & 59.8 & 18 & 1.82 \\
57 & 163.3 & 66.2 & 0.114 & -125 & 165.0 & 66.4 & 25 & 0.71 \\
58 & 140.7 & 67.0 & 0.109 & -95 & 138.6 & 67.0 & 22 & 0.82\\
59 & 145.7 & 37.1 & 0.109 & 15 & 145.1 & 37.9 & 20 & 0.93 \\
60 & 143.4 & 66.6 & 0.106 & -95 & 146.3 & 67.3 & 14 & 1.35\\
61 & 146.3 & 36.5 & 0.105 & -55 & 146.3 & 36.5 & 30 & 0.00\\
62 & 158.6 & 67.2 & 0.104 & -95 & 160.6 & 66.0 & 14 & 1.43 \\
63 & 164.2 & 69.1 & 0.103 & -45 & 164.0 & 69.0 & 28 & 0.12\\
64 & 148.2 & 67.9 & 0.100 & -85 & 146.5 & 67.5 & 23 & 0.75\\
\enddata
\end{deluxetable}

\begin{figure}
\figurenum{1}
\epsscale{.8}
\plotone{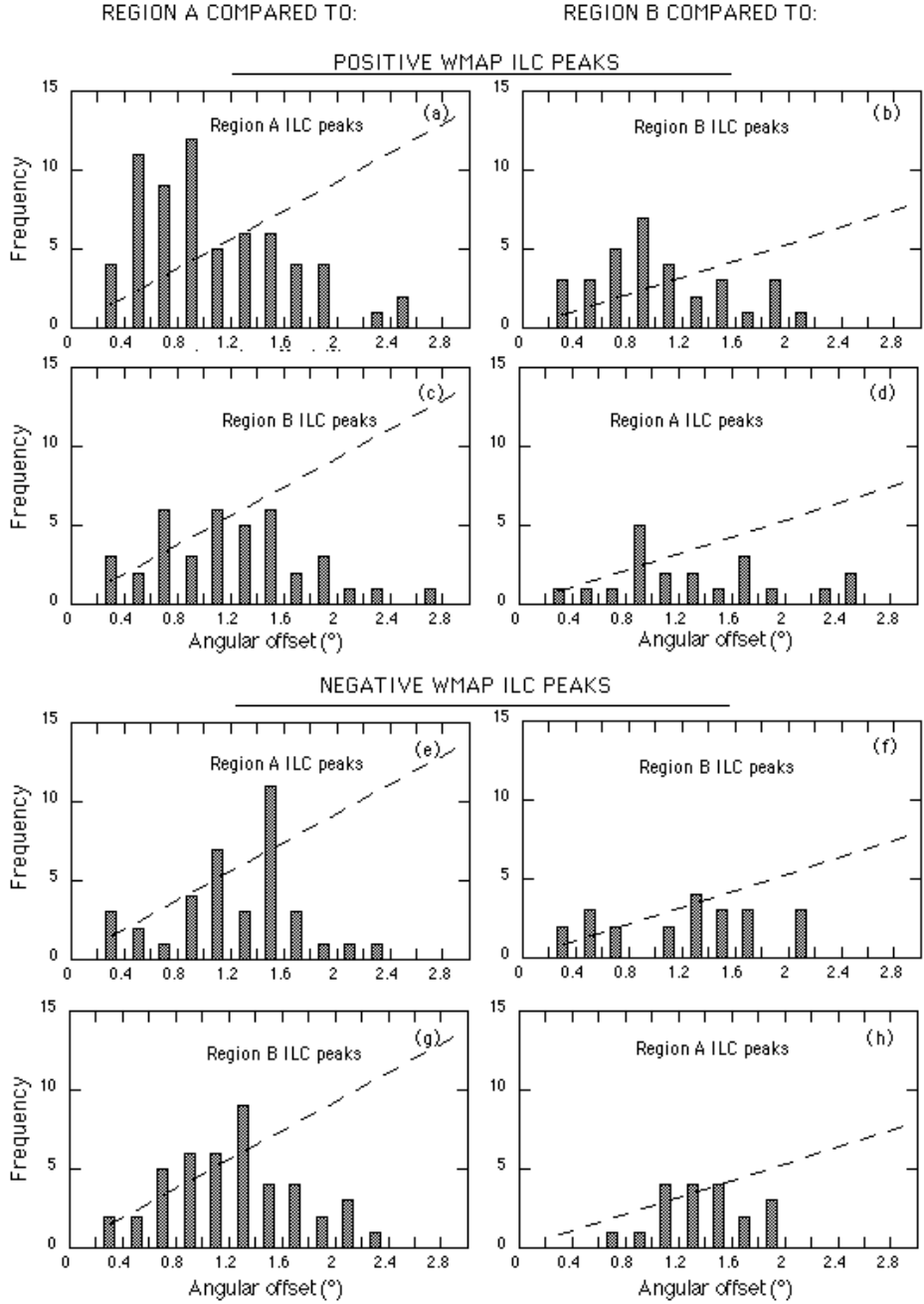}

\caption{The distribution of angular separations between {\it ILC} peaks and closely spaced HI peaks for b ${\geqq}$40\arcdeg for various combinations of {\it ILC} and HI data for the two Regions A \&B, see text.  The dashed line represents a model calculation for a random distribution of peaks according to Eqtn. 1, and the systematic excess of associations with low angular separations with respect to the model in the top two plots supports the hypothesis that the continuum emission arises at the surfaces of HI features.}

\end{figure}

\begin{figure}
\figurenum{2}
\epsscale{.7}
\plotone{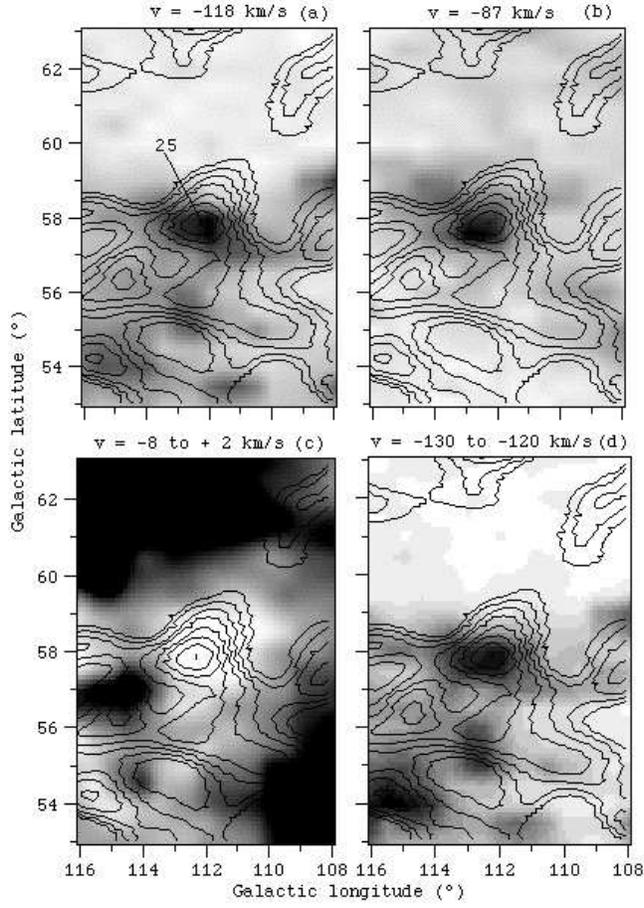}

\caption{ {\it ILC} peak 25 from Table 1 at ({\it l,b}) = (112.\arcdeg 3, 57.\arcdeg 8) at the center of the map (contours from +0.04 in steps of 0.02 mK) overlain on HI (inverted gray scale) structure at the velocities indicated.  In (a) the {\it ILC} peak is associated with a high-velocity HI feature and in (b) with an intermediate-velocity HI feature.  In (c) the {\it ILC} peak is compared to low velocity HI and overlies a clear HI deficit.  It is striking that the integrated high-velocity HI emission displayed in (d) shows two other peaks that are co-located with minima in (c), see text.  These structures strongly suggest that the anomalous velocity gas is interacting with low-velocity HI, producing the {\it ILC} continuum emission in the process.}

\end{figure}

\begin{figure}
\figurenum{3}
\epsscale{.7}
\plotone{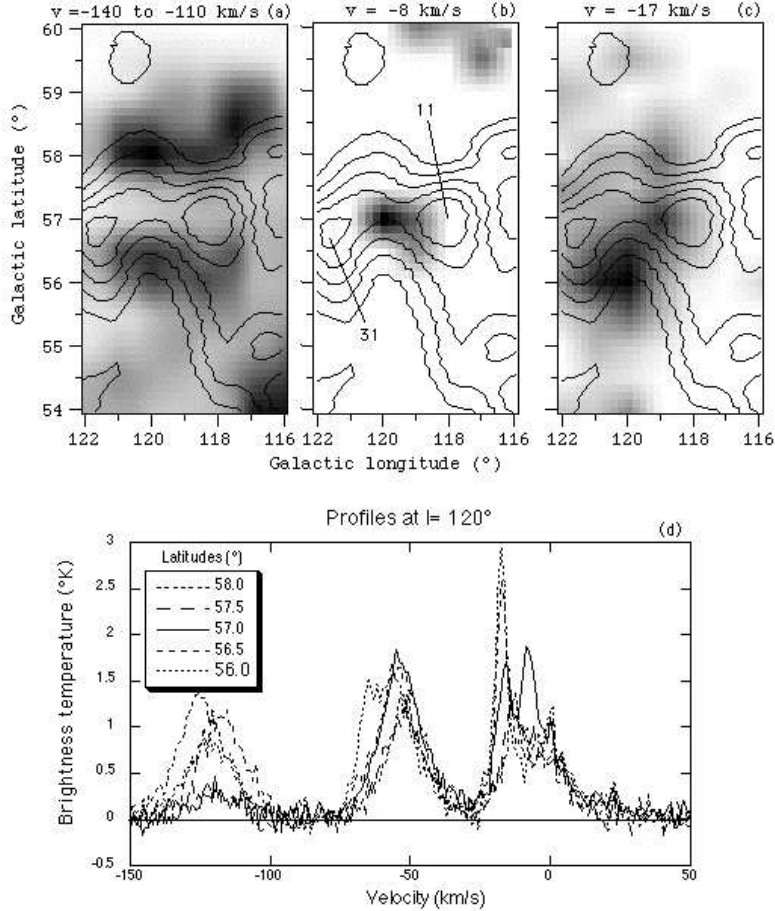}
\caption {(a) The {\it ILC} peaks 11 \& 31 from Table 1 labeled in (b) with contour levels from 0.02 mK in steps of 0.03 mK centered at ({\it l,b}) = (120\arcdeg,57\arcdeg).  A ridge of {\it ILC} structures bridges the gap between the two peaks. In (a) the integrated HI brightness over the velocity range indicated shows two high-velocity features with center velocities determined to be -127 \& -118 km/s (for the north and south components) straddling the {\it ILC} ridge, see text.  Frame (d) shows several HI emission profiles at the latitudes indicated. The HI {\it l,b} map of the striking, unresolved, component at -8 km/s at b = 57\arcdeg\ is shown in (b).  In (c) the area map of the HI component at -17 km/s is compared to the {\it ILC} structure. }
\end{figure}

\begin{figure}
\figurenum{4}
\epsscale{.8}
\plotone{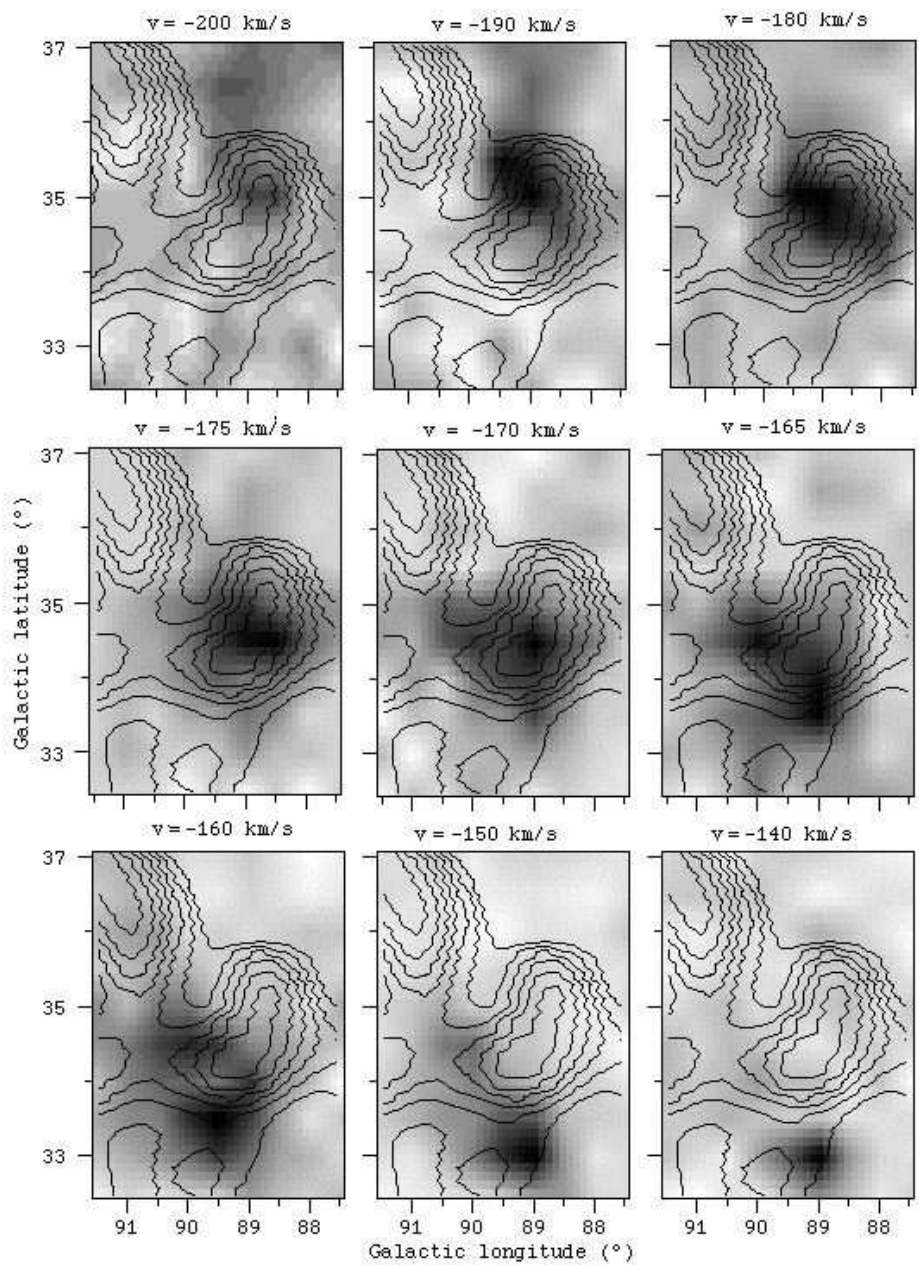}
\caption {The elongated {\it ILC} structure, peak 33, Table 1, at ({\it l,b}) = (89\arcdeg, 34.\arcdeg 5) (contour levels from 0.03 mK in steps of 0.02 mK) showing a direct relationship to the high-velocity HI structure in 1 km/s bands at the velocities indicated.  The HI structure follows the {\it ILC} structure closely as the center velocity shifts.  At -135 and -205 km/s no significant HI emission is detectable.  }
\end{figure}

\begin{figure}
\figurenum{5}
\epsscale{.7}
\plotone{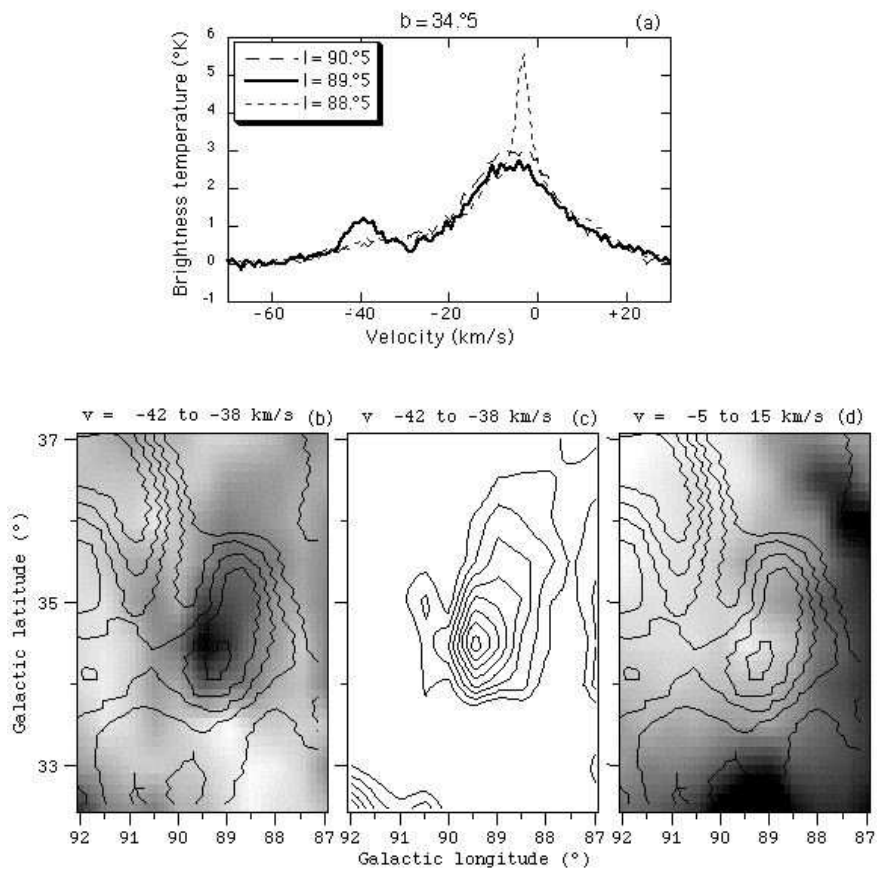}
\caption {Further associations between the {\it ILC} continuum source 33, Table 1, and HI.  (a) HI profiles toward the {\it ILC} peak at the latitude and longitudes indicated.  The profile at longitude 89.\arcdeg5 shows excess emission at -40 km/s.  The gray-scale representation of this feature is shown in (b) with the {\it ILC} contours overlain.  Frame (c) shows the HI data from (b) in contour form to highlight the similarity to the {\it ILC} contours in (b).  In (d) the {\it ILC} peak is overlain on the HI emission at low velocities showing the coincidence with a lack of low velocity emission.}
\end{figure}

\begin{figure}
\figurenum{6}
\epsscale{.8}
\plotone{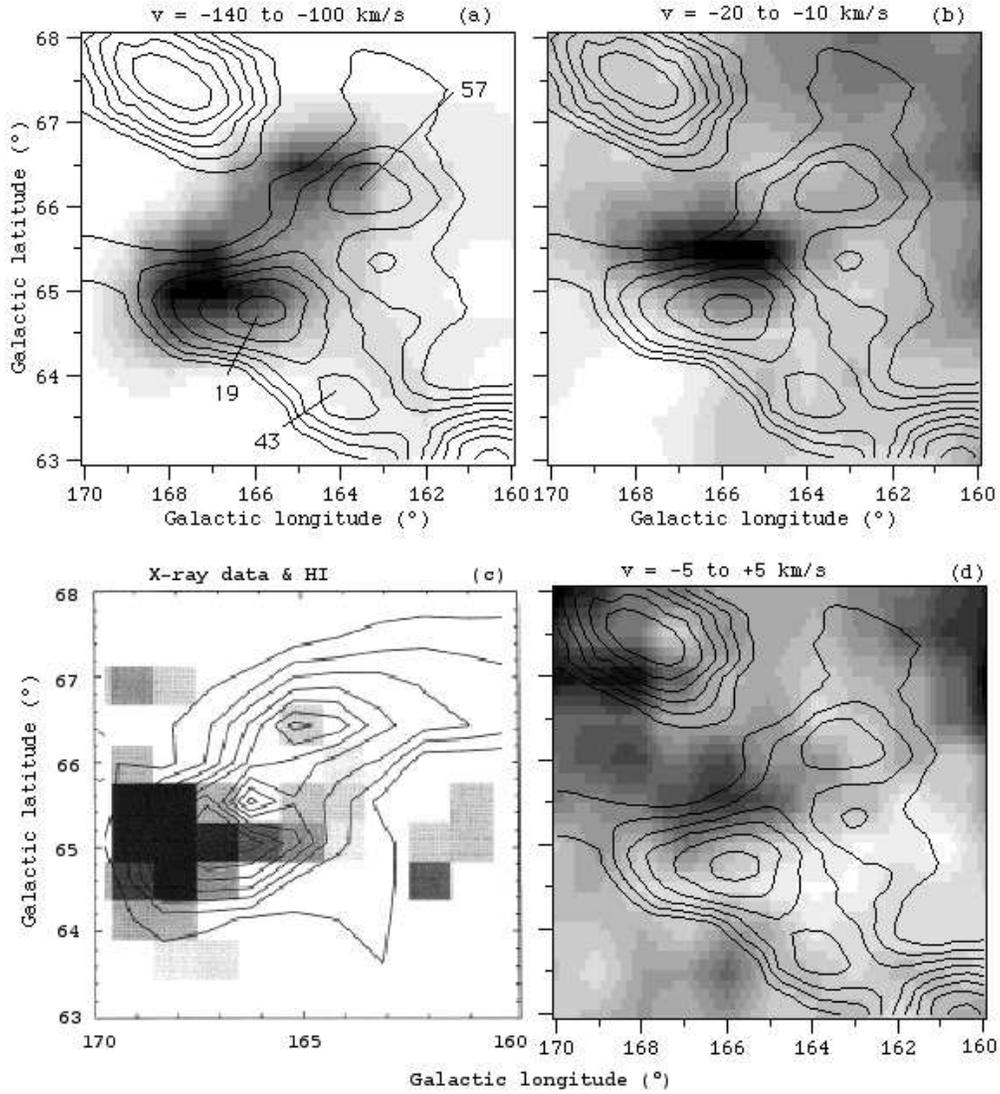}
\caption {{\it ILC} peaks 19, 43 \& 57 around ({\it l,b}) = (165\arcdeg, 65.\arcdeg5) indicated as contours (from 0.06 to 0.20 mK in intervals of 0.02 mK) in (a) (b) \& (d) overlain on inverted gray-scale images of the HI emission toward high-velocity cloud MI at the velocities indicated. (c) The location of a soft X-ray excess emission (pixels) with respect to high-velocity cloud MI contours adapted from Herbstmeier et al. (1995), see text.}
\end{figure}

\end{document}